%% file: Orion_Paper_MNRAS_v5p3.tex
\documentclass[fleqn,useAMS,usenatbib]{mnras}

\usepackage{mathptmx}
\usepackage[T1]{fontenc}
\usepackage{ae,aecompl}

\usepackage{graphicx}
\usepackage[varg]{txfonts}
\usepackage{lscape}
\usepackage{amssymb}
\usepackage{color}
\usepackage{upgreek}
\usepackage[]{units}
\input{library_abkuerzungen}

\usepackage[caption=false]{subfig}
\usepackage{hyperref}	
\hypersetup{colorlinks=true,linkcolor=blue,citecolor=blue,filecolor=blue,urlcolor=blue}

\title[The bimodal initial mass function in the Orion Nebula Cloud]{The bimodal initial mass function in the Orion Nebula Cloud \thanks{Based on observations made with ESO Telescopes at the Paranal Observatory under programme ID 082.C-0032-1}}

\author[H.\,Drass et al.]
{H.\,Drass$^{1,2}$ \thanks{E-mail:\,holger.drass@ruhr-uni-bochum.de},
 M.\,Haas$^{1}$, 
 R.\,Chini$^{1,3}$,
 A.\,Bayo$^{4,5}$, 
\newauthor M.\,Hackstein$^{1}$,
 V.\,Hoffmeister$^{1}$,
 N.\,Godoy$^{4}$, and 
 N.\,Vogt$^{4}$
\\
$^{1}$Astronomisches Institut, Ruhr-Universit\"at Bochum, Universit\"atsstra\ss{}e 150, 44780 Bochum, Germany\\
$^{2}$Department of Electrical Engineering and Center of Astro-Engineering UC, \\
\hspace{0.15cm} Pontificia Universidad Cat\'olica de Chile, Av. Vicu\~na Mackenna 4860, 7820436 Macul, Santiago, Chile\\
$^{3}$Instituto de Astronom\'ia, Universidad Cat\'olica del Norte, Avenida Angamos 0610, Casilla 1280 Antofagasta,Chile\\
$^{4}$Instituto de F\'{\i}sica y Astronom\'{\i}a, Universidad de Valpara\'{\i}so, Av. Gran Breta\~na 1111, Valpara\'{\i}so, Chile\\
$^{5}$Max-Planck Institut f\"ur Astronomie, K\"onigstuhl, D-69117, Germany
}
\date{Received May, 2014; Accepted 2016 May 05}

\pubyear{2016}

\begin{document}

\label{firstpage}

\pagerange{\pageref{firstpage}--\pageref{lastpage}}

\maketitle

\begin{abstract}
  Due to its youth, proximity and richness the Orion Nebula Cloud (ONC) is an ideal testbed to obtain a comprehensive view on the Initial Mass Function (IMF) down to the planetary mass regime. Using the HAWK-I camera at the VLT, we have obtained an unprecedented deep and wide near-infrared $JHK$ mosaic of the ONC (90\% completeness at $K\sim\,19.0\,$mag, $22\arcmin \times 28\arcmin$). Applying the most recent isochrones and accounting for the contamination of background stars and galaxies, we find that ONC's IMF is bimodal with distinct peaks at about 0.25 and $0.025\,\unit{\unit{M_{\sun}}}$ separated by a pronounced dip at the hydrogen burning limit (0.08\,$\unit{M_{\sun}}$), with a depth of about a factor 2--3 below the log-normal distribution. Apart from $\sim$920 low-mass stars ($M < 1.4\,\unit{M_{\sun}}$) the IMF contains $\sim$760 brown dwarf (BD) candidates and $\sim$160 isolated planetary mass object (IPMO) candidates  with $M>0.005\,\unit{M_{\sun}}$, hence about ten times more substellar candidates 
than known before.
  The substellar IMF peak at 0.025\,$\unit{M_{\sun}}$ could be caused by BDs and IPMOs which have been ejected from multiple systems during the early star-formation process or from circumstellar disks.
\end{abstract}

\begin{keywords}
  Stars: brown dwarfs -- Stars: formation -- ISM: dust, extinction -- Infrared: stars
\end{keywords}

\section{Introduction}

  The stellar Initial Mass Function (IMF) \citep{Salpeter55} describes
  the mass spectrum of a stellar population at birth or in young
  star-forming regions shortly after birth. The origin of the IMF is a
  fundamental issue in the study of star formation. Basically, two
  competing theories try to explain the observations: the
  deterministic view postulates that the IMF is essentially determined
  by the Core Mass Function (CMF) in the parental cloud
  \citep{Alves07, Nutter07, Andre10, Hennebelle13}, while the stochastic view
  emphasizes the importance of dynamical interactions and competing
  accretion \citep{Bonnell97, Reipurth01, Vorobyov13}.  

  The high-mass end of the IMF and the peak at intermediate stellar
  masses (0.2\,--\,0.5\,$\unit{M_{\sun}}$) for various star-forming
  regions (e.g. \citet{Bayo11}) support a log-normal IMF shape
  \citep{Miller79, Kroupa01, Chabrier05}. However, at the low-mass end
  the frequency of Brown Dwarfs (BDs) and isolated planetary mass
  objects (IPMOs) is poorly known with large uncertainties
  \citep{Bastian10, Scholz13}. 
  
  The Orion Nebula Cloud (ONC) located at a distance of 414\,pc
  \citep{Menten07} is a benchmark for studying the IMF of young star-forming
  regions and to peer into the substellar mass regime.
  Even so, in the central ($5\arcmin \times  5\arcmin$) region around the Trapezium Star Cluster the bright irregular emission of the nebula and the high extinction hampered the detection of faint objects even at near-infrared (NIR) wavelengths. 
  The age of young stars in the ONC is in the range of 1 -- 5\,Myr; here we adopt 3\,Myr \citep{DaRio11}.

  While in the ONC a growing number of $\sim$60 BDs has been found in recent
  years by means of spectroscopy \citep{Slesnick04, Riddick07, Weights09}, any evidence for a rich BD and IPMO population -- exceeding the log normal extrapolation of the stellar IMF -- is
  still a matter of a debate. On one hand, the IMF appears to be
  steeply declining towards the low-mass (BD) end
  \citep{Hillenbrand00, DaRio11}. In these cases, however, the
  reported IMF decline runs below the total number of spectroscopically confirmed BDs, combined from the authors mentioned above. This indicates a problem in these photometric searches. On the other hand, \citet{Muench02} have claimed
  an upturn of the IMF in the BD mass regime, albeit at their
  observational brightness limits and therefore quite speculative.
  Finally, \citet{Lucas05} reported on the so far most 
  robust IMF study of the ONC, which reveals a dip with 
  potential rise of the low-mass IMF or simply a broad IMF plateau; 
  as pointed out by \citet{Lucas00}, imperfections in the available isochrones 
  prevent to easily distinguish between the two IMF shapes. Optical wide field
  studies covering the outer ($30\arcmin \times 30\arcmin$) ONC
  regions suffer from extinction and failed to reveal a rich
  BD population \citep{DaRio11}. Still, the substellar IMF in Orion is
  controversial and a comprehensive investigation employing deeper
  data and state-of-the-art isochrones is needed.   
  
  \section{Data}
\begin{figure}
   \centering
   \includegraphics[width=\hsize,keepaspectratio, angle=0, trim=3cm 2cm 3cm 3cm, clip=true] {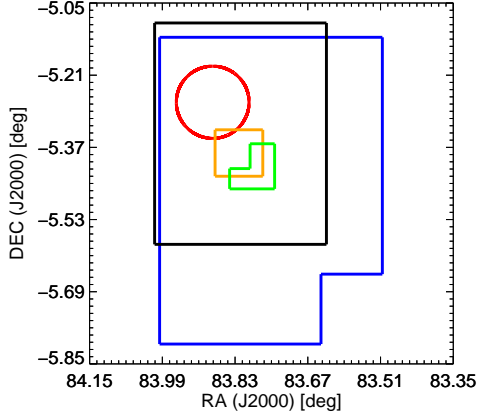}
   \caption{Surveyed areas covered by different studies, in black the area covered by HAWK-I used in this work, in blue the area from \citet{Robberto2010}, in orange the data from \citet{Muench02}, and in green the data from \citet{Lucas05}. The circle of 10' diameter marks the region around M43.}
   \label{Fig:FoVs}
  \end{figure}

  Using the HAWK-I camera at the VLT on November 8--11, 2008, we
  obtained a large $22\arcmin \times 28\arcmin$ and deep image mosaic
  of the ONC in the $JHK$ filters centred at $1.25$, $1.65$ and $2.15\,\upmu$m (i.e. $K_{s}$ here for short denoted $K$)
  under good seeing conditions (FWHM $<0.7\arcsec$).
  The central mosaic position is at  RA $= 05^\textrm{h} 35^\textrm{m} 16^\textrm{s}.68 $ and  Dec $= -05^\circ 20'22\farcs2 \left(\textrm{J}2000\right) $. 
  In Fig.\,\ref{Fig:FoVs}, the area surveyed by HAWK-I is displayed together with the field of views of the comparison data from \citet{Robberto2010,Muench02}, and \citet{Lucas05}.
  All fields were observed ten times with a standard dither pattern in $J,H,K$ and $1\,s$ exposure time per frame.

  \subsection{Basic reduction and distortion correction}
    The basic data reduction was performed with IRAF.\footnote{The Image Reduction and Analysis Facility (IRAF)(ver.\,2.14.1) is distributed by the National Optical Astronomy Observatories, which are operated by the Association of Universities for Research in Astronomy, Inc., under cooperative agreement with the National Science Foundation.}
    For calibration we took regular dark and flat exposures close to the observations and combined them into  master dark and  master flat. Subtraction of the master dark and division by the master flat yielded a master sky frame that was subtracted from the science frames.

    At the time of the observations there was no suitable distortion map available. To achieve a solution for the distortion correction the IRAF/MSCRED package was applied.
    After distortion correction, the comparison with 2\,MASS \citep{2mass} showed that for objects with high quality detection according to the 2\,MASS point source catalog (those labelled `AAA') the differences are within 0.2" in 98\% of the sources.

  \subsection{Instrumental magnitudes}
    All photometric operations were performed on the 144 quadrant images separately using the IRAF/DAOPHOT package. A preliminary source list was created with IRAF/DAOFIND. Due to the strong and variable nebular emission across the field the parameters were adjusted in a way to register only sources with a high signal-to-noise ratio (S/N $>50$).
    False detections such as halo detections around saturated stars and saturated stars themselves ($K\,<\,12.5$\,mag) were omitted. In addition, the positions of all point like structures where added manually using IRAF/DAOFIND interactively. Then, in a preparation step and for later comparison, aperture photometry with IRAF/PHOT was done. To perform point spread function (PSF) photometry of all sources we selected the most isolated, non saturated stars in each quadrant with IRAF/PSTSELECT and created a PSF with IRAF/PSF. The routine IRAF/ALLSTAR calculated the PSF magnitudes for all sources and subtracted their PSF from the image.
    The resulting image was inspected by eye and newly appearing objects, for instance faint close companions to a bright star, were added to the source list. After repeating the PSF photometry and source subtraction process the images were reinvestigated. This process was repeated until all sources were removed from the frames.
    In order to receive sharpness and roundness for the sources added by hand IRAF/DAOFIND was used with a very low threshold (S/N $=0.5$). Then the coordinates where transferred from image coordinates (xy) to the world coordinate system (WCS) (RA, Dec) employing the distortion corrected image header by using the WCSTOOLS/xy2RaDec \citep{wcstools} task. To find a photometry for all sources the ``handmade'' source list was matched with the low threshold sources utilizing the ALADIN/xmatch task.

  \subsection{Photometric calibration with 2 MASS}
    The absolute calibration was done by comparison with 2\,MASS. Each detector quadrant was calibrated separately and only the objects with the best quality flags `AAA' fainter than $12$\,mag where used. First the average difference and the mean deviation for the difference between the magnitudes from 2\,MASS and HAWK-I were calculated.
    This procedure yielded a couple of sources which deviated significantly from the mean standard deviation.
    Visual inspection of these cases showed that the deviation is well explained by the reduced spatial resolution of 2\,MASS compared to \mbox{HAWK-I} and the increased photometric errors in nebular regions. These outliers have been excluded from the calibration. The remaining differences were averaged a second time and added to all instrumental magnitudes.    

\begin{figure}
      \includegraphics[width=\hsize,keepaspectratio, angle=0, trim=1cm 3cm 2cm 4cm, clip=true]{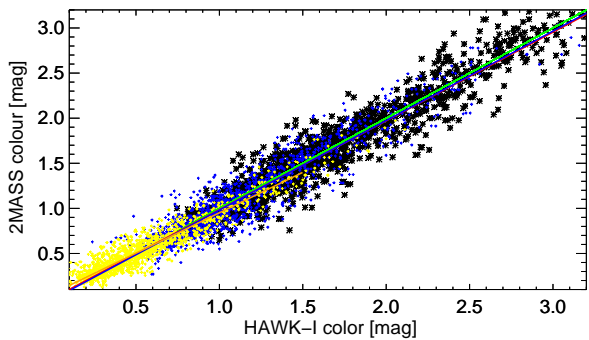}
      \caption{Investigation of a color term correction for the 2\,MASS and HAWK-I filter combinations. The green line's slope is 1 and intersects at zero. The red, blue, orange line marks the fitted solution for $(J-K)$,$(J-H)$, $(H-K)$, respectively.The black, blue, yellow crosses show the colors for $(J-K)$, $(J-H)$, $(H-K)$, respectively.}
      \label{(J-K)_2m_vs_(J-K)_HAWK-I}
    \end{figure}

    Additionally, since 2\,MASS and HAWK-I are using different filter systems, we investigated the color term.
    The comparison between the colors  $(J-K)$, $(J-H)$, $(H-K)$ for HAWK-I and 2\,MASS are plotted in Fig.~\ref{(J-K)_2m_vs_(J-K)_HAWK-I}.
    \newpage
    The results of a linear fits are given in the following equations.
    \begin{eqnarray*}\label{eq:lin_fit}
    (J-Ks)_{2\,MASS}= (0.98 \pm 0.01)~(J-K{\rm s})_{\rm HAWK-I} - (0.034 \pm 0.004)\nonumber\\
    (J-H)_{2\,MASS}= (0.99 \pm 0.01)~(J-H)_{\rm HAWK-I} - (0.03 \pm 0.02)\\
    (H-Ks)_{2\,MASS}= (0.87 \pm 0.03)~(H-K{\rm s})_{\rm HAWK-I} + (0.063 \pm 0.002)\nonumber
    \end{eqnarray*}
    \begin{figure*}
      \centering
     \includegraphics[width=\hsize, angle=0, trim=0cm 0.0cm 2cm 17cm, clip=true]{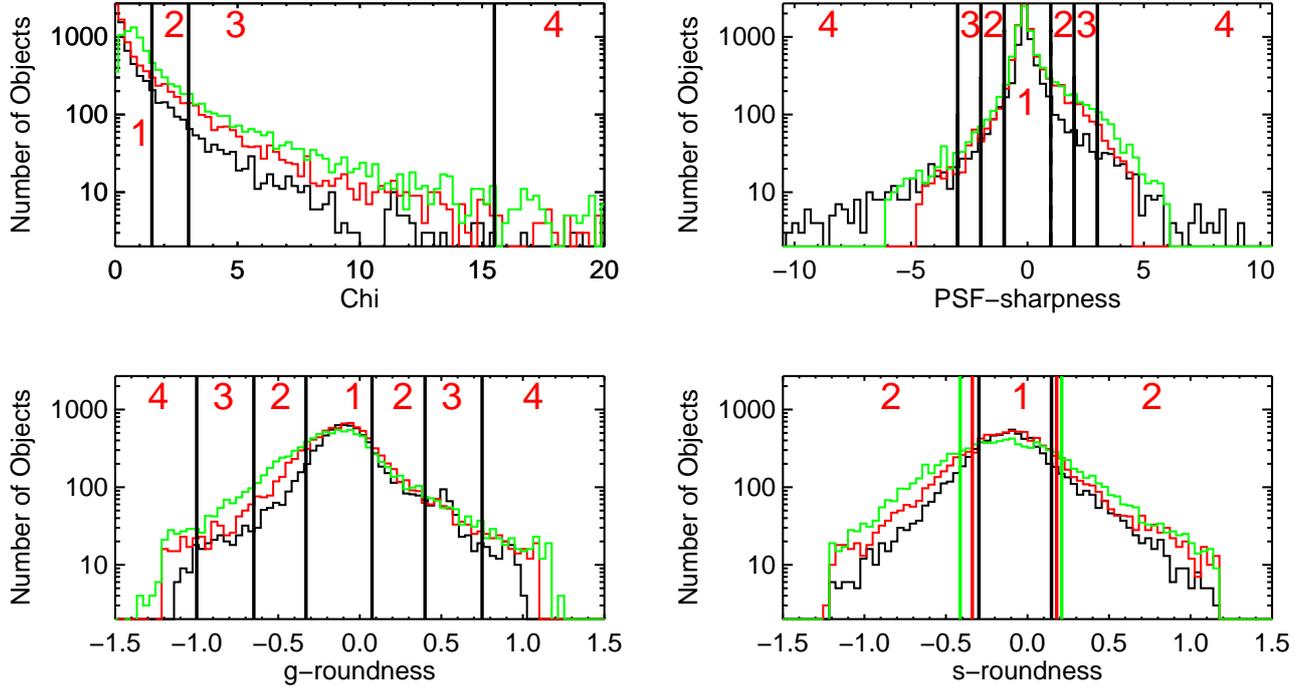}
      \caption{Quality flag plots. For the chi, PSF-sharpness, g-roundness the numbers give the flag value in shown range. For the s-roundness the limits are filter dependent. (Black, red, green are the average plus the mean standard deviation for $J$, $H$, and $K$, respectively.)}
      \label{Quality_flag_plots}
    \end{figure*}
    Most of our objects have colors in the range $0.7 < (J-K) < 3$ resulting in negligible color corrections.
    Given that the red colors of these sources are dominated by extinction we expect that possible uncertainties in the extinction law are much larger than this filter term. The results for $(J-H)$ and $(H-K)$ are similar. In summary, we did not apply any color term correction.

  \subsection{Source selection}
    The resulting catalogs of all calibrated objects in $J$, $H$, and $K$ contain 7975, 9630, and 9899 sources, respectively. To exclude any false detections like artefacts or poorly determined sources, a refined selection based on the search and photometry parameters was performed. In a first step the sources were conservatively selected by their sharpness and roundness (ratio of best fitting Gaussians) derived from aperture photometry and the sharpness calculated during the PSF photometry by having a deviation less than five times the mean absolute deviation from the corresponding average. By eye inspection of the rejected sources showed that they were either saturated or extremely faint; other rejected sources were simply observational artefacts in the frame.
    For a second selection level the magnitude errors were fitted and quality flags (QF) for the fit (chi), s-roundness, g-roundness, PSF-sharpness and magnitude error were created. The corresponding values for the quality flags are plotted in Fig.~\ref{Quality_flag_plots}.
    Sources with the following properties were rejected:
    \begin{itemize}
      \item brighter than  $12.5$\,mag in all filters
      \item magnitude errors at the object brightness larger than ten times the fitted average magnitude error
      \item magnitude errors greater than $0.1$\,mag
      \item quality flag worse than 3 for the PSF- sharpness
    \end{itemize}
    By analyzing the different quality flags it turned out that the combination of several flags helps to find bona fide objects. Therefore the sum of the quality flags for chi, s-roundness, g-roundness, PSF-sharpness and magnitude error was calculated and only sources with an empirically determined value less \mbox{than $13$} were accepted.

  \subsection{Overlap handling}
    The mosaic was arranged with a sufficient overlap between the adjacent fields. Objects located in overlapping regions were thus measured up to four times. In order to obtain only one measurement per source each identification in a radius of $0\farcs4$ was recorded. For those objects that only have a second record, the difference between the two magnitudes is calculated. If the difference is smaller than 0.15\,mag or smaller than ten times the fitted mean value of all objects in the magnitude bin $\Delta m = 0.5$\,mag, the mean value of these two identifications is used. Otherwise the quality flags for the error and the chi value of both records are evaluated. If both quality flags have a value of $4$, the identifications are rejected. In a last step the distance to the edge of the image frame is analyzed. If both identifications are close to an edge, the object is also rejected. If one record is close to an edge we used only the record of the other identification. In this way, all questionable identifications for objects with two records could be handled.

    For objects with three records the magnitude difference between all three identifications was first considered with the same criteria as above. If all three identifications fulfil the conditions, their average is used as the best value. If not, the record with the worst magnitude error is rejected and the criteria are reconsidered. In case of acceptance, the mean is evaluated and used as the brightness of this object. The remaining identifications are then investigated for their relative distance from the border as in the case of two records. Also acceptance and rejection is done in the same way. This solved the ambiguities for all objects.

    Finally, when an object was recorded four times the average is calculated and only those two identifications with the smallest difference relative to the average were considered further on. Then the difference between these remaining values was considered in the same way as above. We found that no further selection criteria were necessary. For these objects again the mean of the measurements was used as the final magnitude.
    
    The resulting master catalog contains all reliable sources fainter than $12.5$\,mag in all filters contained in our HAWK-I FOV; brighter sources are not in the linear range of the detector and thus omitted. To construct a fairly complete catalog of ONC members we added brighter sources from the list by \citet{Robberto2010} covering nearly the same area; this list is a compilation including data from \citet{Muench02} and the 2\,MASS archive. 
    
\begin{figure}
     \centering
     \includegraphics[width=\hsize,keepaspectratio, angle=0, trim=0cm 0cm 0cm 0cm, clip=true] {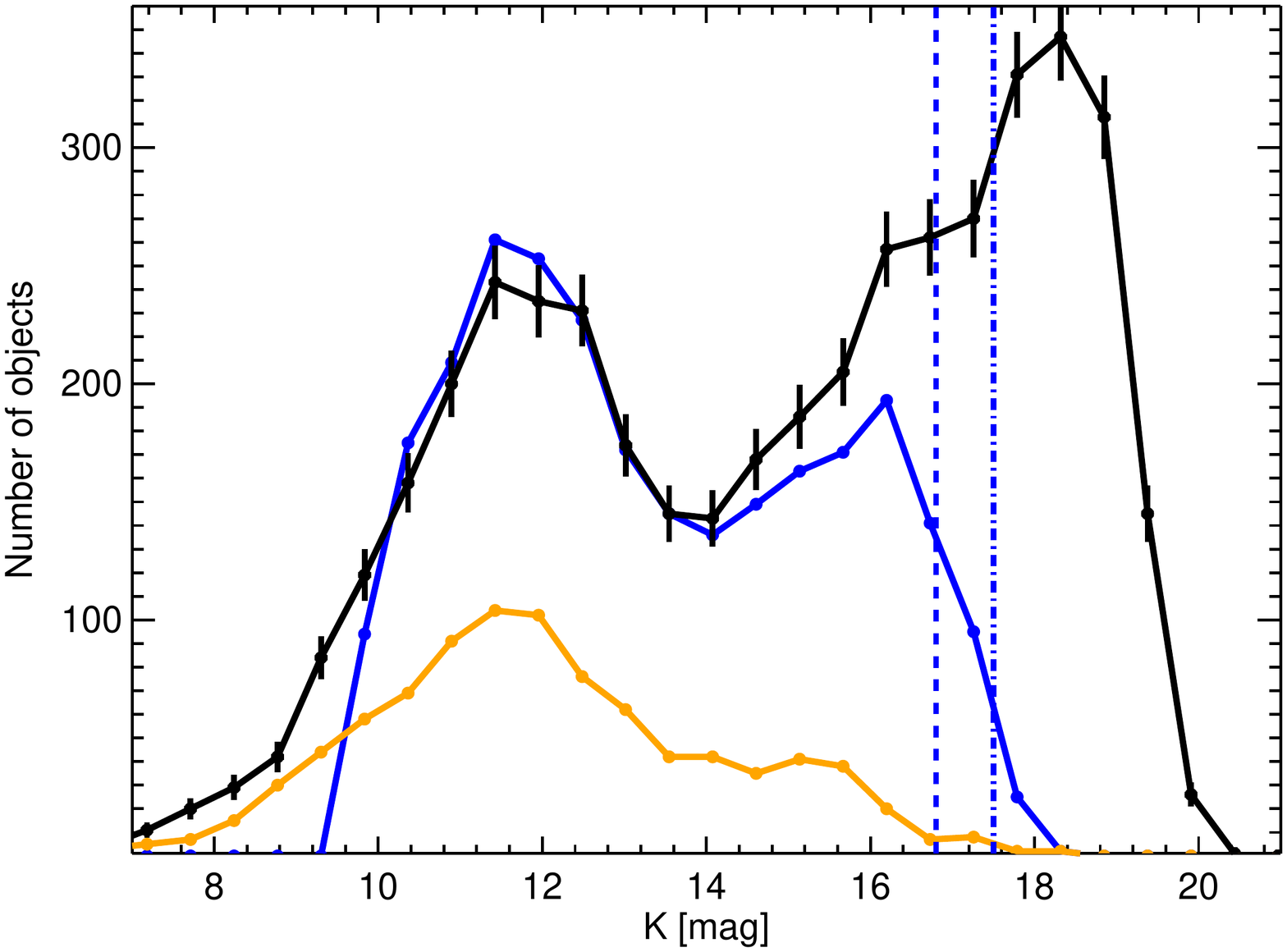}
     \caption{$K$-band luminosity functions for the HAWK-I data (black) compared to \citet{Robberto2010} (blue) and \citet{Muench02} (yellow) without completeness correction applied. The error bars correspond to $\sqrt {N}$. The vertical dashed and dash dotted lines mark the completeness limits of \citet{Robberto2010} at 90\% and 70\%, respectively.}
     \label{Comparison_Rob}
    \end{figure}

    Hence, both catalogs (HAWK-I and Robberto's) overlap by $K = 4.5$\,mag. The same overlap range of about $4.5$\,mag holds for $J$ and $H$. The comparison presented in Fig.\,\ref{Comparison_Rob} demonstrates that the HAWK-I data reach 2\,mag deeper than the previous data, which start to suffer from incompleteness in the BD regime (vertical lines in Fig.\,\ref{Comparison_Rob}).
    The total number of sources in the added catalog is $4340$. 
    All photometric errors are smaller than $0.1\,$mag in all three filters. Thus, all three bands can be used to infer the object's extinction and mass from color-magnitude diagrams (CMDs).
    Similar to \citet{DaRio11}, from here on, we excluded a circle with 10$\arcmin$ diameter centred on the nebula M43, because it is a distinct cluster with its own, possibly different, mass function. 
    The HAWK-I FoV is almost covered by the data from \cite{Robberto2010}. For a consistent literature comparison only sources in the common field of view are further analysed.
    The resulting catalog for all sources measured in all three filter is given in Table~\ref{member_cat}.

  \subsection{Completeness}\label{completeness}
\begin{figure*}
      \includegraphics[width=\hsize, angle=0, trim=1cm 0cm 0cm 0cm, clip=true]{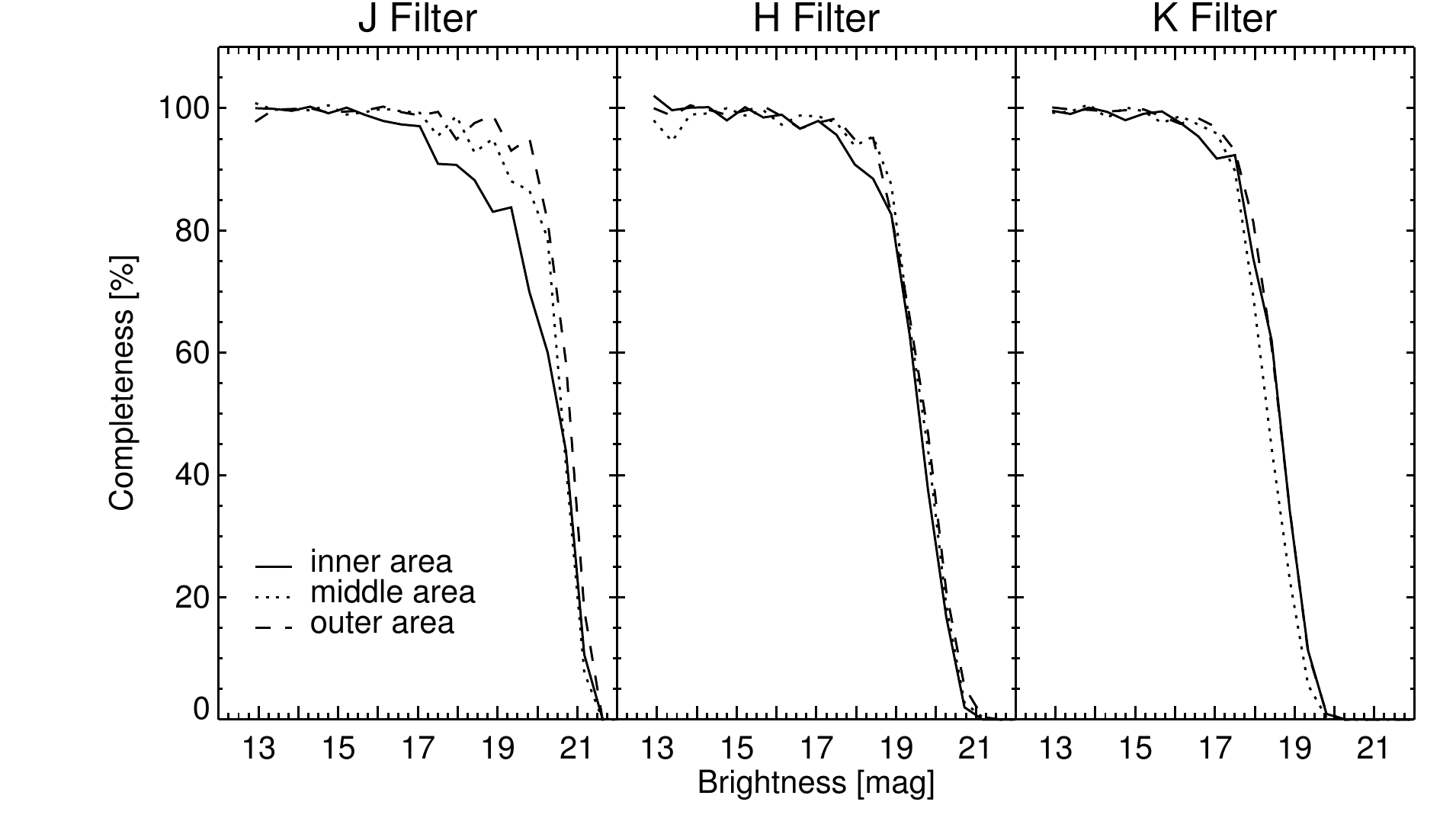}
      \caption{Completeness function for $JHK$ in the inner (\mbox{radius $(r) < 4'$}, solid curve), intermediate ($4'< r < 8'$, dotted curve) and outer region ($8'< r < 12'$, dashed curve). Excluded is the area of $5\arcmin$ radius around the nebula M43.}    
      \label{fig:Compleness_JHK}
    \end{figure*}

    To determine the completeness of our data set we used the routine IRAF/ADDSTAR to randomly add artificial stars down to 22\,mag.    
    Than we ran IRAF/DAOFIND with a $0.1 \sigma$ threshold, which -- according to our tests -- corresponds to the by eye selection for the real sources.
    For the photometry IRAF/PHOT and IRAF/ALLSTAR were utilized. The $xy-$coordinates of the artificial stars and detected candidates were transformed to RA and Dec by using WCSTOOLS/xy2sky.
    In order to find only the artificial stars a cross match using ALADIN/XMATCH with the original list of artificial stars was performed.
    To ensure that only well measured objects were re-discovered only objects with an error smaller than 0.1\,mag and a difference between injected and re-discovered object smaller than 0.5\,mag were accepted. In Fig.~\ref{fig:Compleness_JHK} the completeness obtained from the artificial star experiment is shown; the $90\%$ completeness limits are listed in Table\,\ref{Tab:completeness}.
    \begin{table}\centering
      \begin{tabular}{c|c|c|c}
        Filter & inner area  & middle area & outer area\\\hline
        $J$ & 18.0 & 19.5 & 20.0 \\\hline
        $H$ & 18.0 & 18.5 & 18.5 \\\hline
        $K$ & 17.5 & 17.5 & 17.5 \\
      \end{tabular}
      \caption{$90\%$ completeness limits in mag.}
      \label{Tab:completeness}
    \end{table}

   \subsection{Color-Color Diagram}
\begin{figure}
        \includegraphics[width=\hsize,keepaspectratio, angle=0, trim=0cm 0cm 0cm 0cm, clip=false]{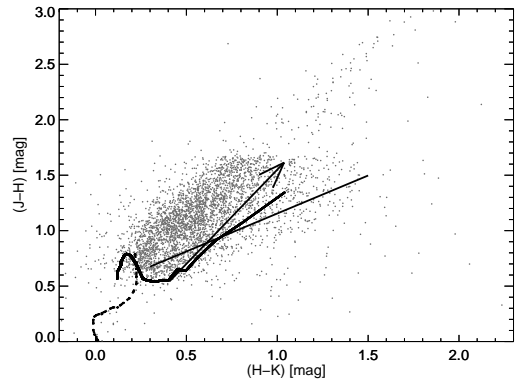}
        \caption{Color-Color Diagram for sources measured in three filters (grey dots), excluded are sources around M43. 
        The black dash-dotted curve is the main sequence from \citet{Ducati2001} while the black solid curve indicates the 3\,Myr isochrone from \citet{Baraffe15} and \citet{Allard13}. The straight line is the locus of the classical TTauri stars \citep{Meyer97}; the reddening vector for $A_V = 10$ is also shown \citep{Rieke85}.}
        \label{CCD_JmH_vs_HmK}
    \end{figure}

    Fig.~\ref{CCD_JmH_vs_HmK} displays the color-color-diagram for sources measured in three filters, excluded are sources around M43 as mentioned before. 
    The color-color-diagram shows that only a small fraction ($<$\,5\%, 195 sources) of objects lie to the right of the isochrone. Assuming they are Orion members, they exhibit a $K$-band excess from circumstellar material. Beside objects exhibiting $K$-band excess also objects without excess can be members. Actually, most objects do not show a $K$-band excess.
    \footnote{
    A preliminary analysis of Spitzer/IRAC photometry (3.6\,$\mu$m and 4.5\,$\mu$m) shows IR excess for 30\% of the objects brighter than $K=13$\,mag. 
    This is consistent with an age of about 3\,Myr. Because of the low irradiation power no excess at shorter wavelength can be expected.
    For fainter objects the IRAC photometry becomes incomplete.}

\section{The observed color-magnitude diagrams}

\begin{figure*}
      \centering
      \subfloat[Color-magnitude diagram $K$ vs $J-K$ of the \mbox{HAWK-I} data (black dots). The thick blue, red, black and orange curves show the 1\,Myr, 2\,Myr, 3\,Myr and 5\,Myr isochrones \citep{Baraffe15,Allard13}, respectively. All isochrones are shown for masses between 1.4\,$\unit{M_{\sun}}$ and 0.003\,$\unit{M_{\sun}}$. The arrows are extinction vectors of length $A_V = 10\,$mag starting from the isochrone at the upper (0.08\,$\unit{M_{\sun}}$) and lower (0.012\,$\unit{M_{\sun}}$) mass limit of brown dwarfs. The dashed line at $J-K<2.6\,$mag presents the sample limitation at $A_V < 10\,$mag.]
      {\includegraphics[width=.48\linewidth, keepaspectratio, angle=0, trim=0cm 0cm 0cm 0cm, clip=true]{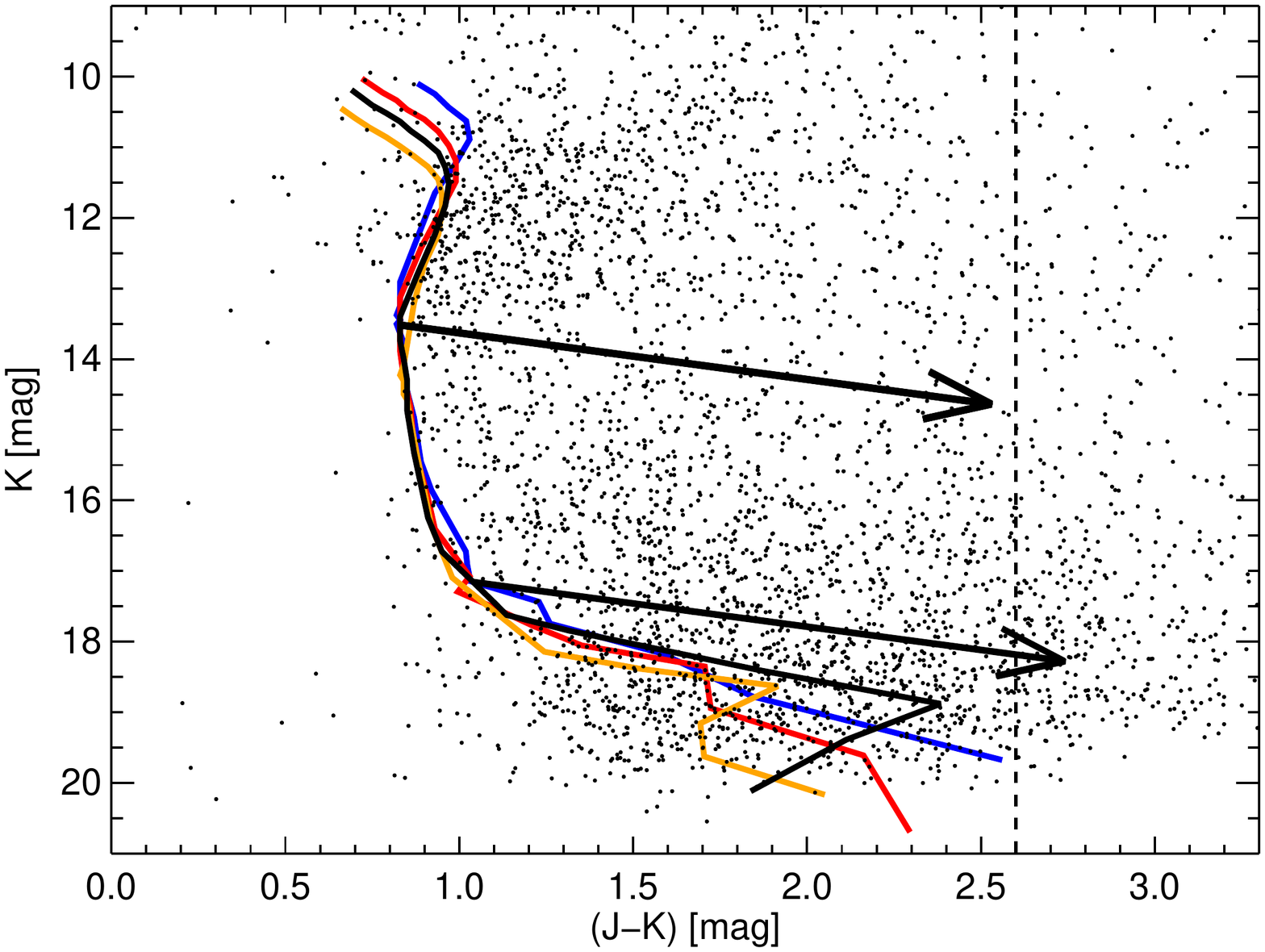}\label{Fig:CMD_data}} \quad
      \subfloat[Color-magnitude diagram $K$ vs $J-K$ of the known spectroscopically verified brown dwarfs (red dots) and background stars from the Besan\c{c}on model (yellow dots) and background galaxies from UKIDSS ultra-deep field (blue crosses). The CMD position of the background objects is shown without reddening by the Orion dust screen. The rest of the symbols is the same as in the left diagram.]{\includegraphics[width=.48\linewidth,keepaspectratio, angle=0, trim=0cm 0cm 0cm 0cm, clip=true]{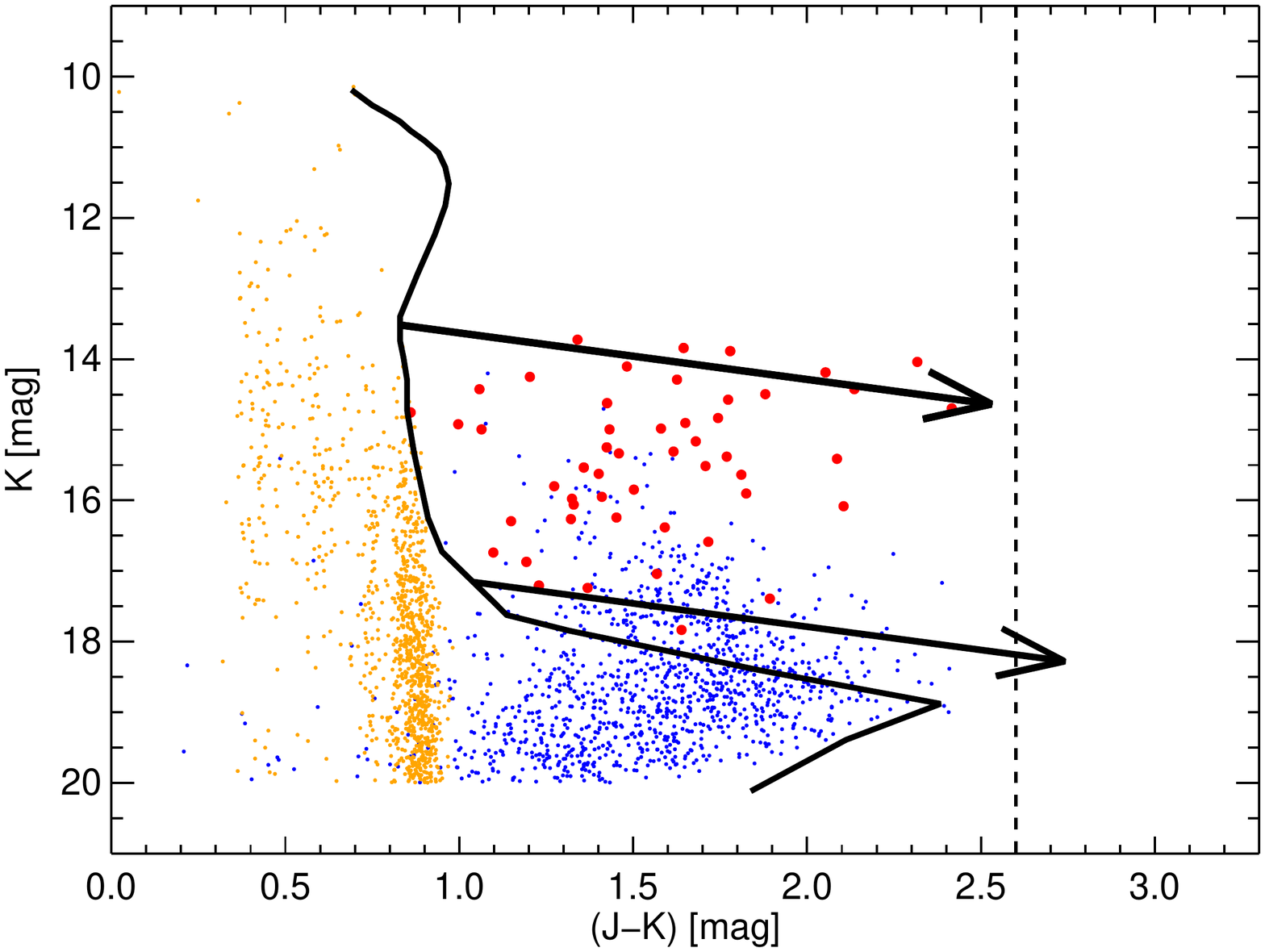}\label{Fig:CMD_bck_KBD}}
      \caption{Color-magnitude diagrams}
     \label{fig:CMDs}
  \end{figure*}

  Figure\,\ref{Fig:CMD_data} presents the $K / (J-K)$ CMD of the observed sample.
  The 3\,Myr isochrone from \citet{Baraffe15} holds for a mass range of 1.4\,--\,0.01\,$\unit{M_{\sun}}$. To cover the full substellar range as far as possible, the 3\,Myr isochrone from \cite{Allard13}
  \footnote{\cite{Allard13}, CIFIST2011bc -- Model, 2\,MASS filter set, Vega magnitudes, see appendix \ref{App:Allard_3Myr_isochrone}} 
  was used to extend the mass range to 0.003\,$\unit{M_{\sun}}$ . Allard's isochrone accounts for NIR spectral features of substellar objects (e.g. \citet{Canty13}), which were not taken into account in former isochrones.
    
  We display the $A_V = 10\,$mag extinction vectors at the hydrogen and deuterium burning limits adopting the standard
  galactic extinction law \citep{Rieke85}. 
  
  Compared to previous CMDs of the ONC \citep{Hillenbrand00, Muench02, Lucas05} the \mbox{HAWK-I} data show
  between 5 to 10 times more sources.
  
  Known spectroscopically verified BDs \citep{Slesnick04, Riddick07, Weights09} lie essentially between these boundaries (red dots in Fig.\,\ref{Fig:CMD_bck_KBD}). 
  The few BDs outside the
  boundaries may be explained by age spread (1\,--\,5\,Myr vs. the 3\,Myr adopted). The younger/older isochrones virtually coincide
  with the 3\,Myr isochrone, except that for a given mass the object's
  brightness declines with age; but the differences of the 1 and 
  5\,Myr isochrones compared to the 3\,Myr isochrone are small
  ($\Delta K<0.5\,$mag) within the 3\,Myr BD range  
  ($13.8\,$mag$\,<\,K\,<\,17.2\,$mag). While the isochrones by \citet{Baraffe15} is being extensively tested against observations in the brown dwarf regime, the isochrone from \citet{Allard13} may be somewhat uncertain in the mass range of IPMOs as indicated by the distinct CMD structures at $K > 18.5\,$mag and $J-K \sim 1.5$ calling for some reservation  when constructing the IPMO mass function. 

  A striking feature of the CMD (Fig. \ref{Fig:CMD_data})
  is the low number of sources along the stripe parallel to the $A_V$
  vector starting at the upper BD mass boundary of the isochrone
  (M\,=\,0.08\,$\unit{M_{\sun}}$, at $K=13.8\,$mag), compared to the stripes
  above and below that line. 
  Visual inspection already indicates a deficiency of sources close to the hydrogen burning limit and a rise of the number of potential substellar sources. The same holds for
  all other color-magnitude combinations ($J/(J-H), J/(J-K), J/(H-K)$, etc.). 
  The large overlap in magnitude and the good agreement between the \mbox{HAWK-I} and the \cite{Robberto2010} catalogs (Fig.\,\ref{Comparison_Rob}) support the idea that the deficiency-stripe is not an artefact caused by the merge of the two catalogs.

  At the distance of Orion (414\,pc), contamination of our sample by
  foreground objects unrelated to the Orion complex is negligible; less than 150  objects are expected from the galactic model of \citet{Robin03}. Regarding populations related to the Orion complex,
  \cite{Alves12} and \cite{Bouy14} found 2123 foreground candidate members in about 10 square degrees towards Orion including the clusters NGC\,1980 and NGC\,1981. After comparing with the HAWK-I mosaic, we have identified and excluded 45 sources belonging to the foreground cluster. 
  
  Background stars and galaxies are expected to influence the source counts significantly.
  For the following analysis we are using an extinction and magnitude limited sample.
  The data are limited to $J-K<2.6\,$mag, to obtain an extinction limited sample ($A_V < 10\,$mag). While this cut leads to some incompleteness for ONC members fainter than $K = 18.5\,$mag and  $J-K>2.6$, i.e. IPMOs below 0.006\,$\unit{M_{\sun}}$, it enables a proper estimate of the contamination by background stars from the Besan\c{c}on model and background galaxies from UKIDSS ultra-deep field in the Brown dwarf mass regime.
  This is already a conservative assumption, since in \citet{Shimajiri2011} and \citet{Ripple13} 
  the extinction maps show that the extinction in Orion can be much larger.
  
  \section{Accounting for background objects} \label{Sec:Accounting_for_bck_obj} 
  Since the targeted sources are very faint, so far, there is no membership criterion for each source available. Instead, we take advantage of the fact that the background sources suffer heavily from
  extinction by the Orion nebula cloud and use this fact to correct for contamination of the IMF by background objects.
  To determine the mass function, the sources are shifted (dereddened) in the CMD to the isochrone along the direction of the AV vector.
  Note that this dereddening changes the resulting luminosity functions (LFs), because highly reddened source now become brighter.
  
  At each position of the HAWK-I map, the total extinction $A_V$ through the cloud can be determined from CO observations. We used the $^{12}$CO map with a $7.5\arcsec$ pixel size \citep{Shimajiri2011} and the $^{13}$CO map with $20\arcsec$ pixel size \citep{Ripple13} and produced two extinction maps with consistent results.  
  
  Our HAWK-I data constitute an extinction limited sample with $A_V<10\,$mag ($J-K<2.6$). For about $23\%$ of
  the HAWK-I mosaic area the extinction map shows $A_V<10\,$mag which means that only $0.04$ square degrees are transparent for background objects. The background contamination consists of two components, stars and galaxies.
   
  We used the Besan\c{c}on model of our Galaxy \citep{Robin03} to create a sample of background stars in a cone of $0.04$ square degree and a distance larger than $400$\,pc in the direction of Orion (at galactic latitude $-19.4\deg$). Including all spectral types and Galaxy components in the model parameters we obtain about $860$ background stars. Fig.\,\ref{Fig:CMD_bck_KBD} shows the location of the predicted background stars in the $K/(J-K)$ CMD. Note that in Fig.\,\ref{Fig:CMD_data} the total observed population is naturally reddened by the molecular cloud, but the model predictions for background stars (yellow dots in Fig.\,\ref{Fig:CMD_bck_KBD}) do not yet include any reddening by the Orion nebula cloud. As a consequence the unreddened background stars are located in the CMD at their intrinsic colors. The few stars at $J-K<0.8$ play a minor role. The bulk of the stars lies at $0.8<J-K<1.0$ with a median at $J-K=0.89$. In practice, due to reddening by the ONC, they would be moderately reddened ($A_V<10\,$mag) and thus shifted along the direction of the $A_V$ vector. If they were reddened by more than $A_V\sim10\,$mag, the majority of the background sources would be shifted beyond our extinction limit ($J-K=2.6$). The same applies for other CMD combinations (e.g. $H/(H-K)$ etc.). To reach our aim to find the LF of the ONC, the contaminating background stars need to be subtracted from the total LF. Therefore, each star is moved along the reddening vector until it reaches the isochrone and the magnitude of the intersection is assigned to it.
    
  To estimate the contribution from background galaxies, we use the UKIDSS ultra-deep $J$ and $K$ data set of the ELAIS-north field, a region at high galactic latitude with very low extinction. Fig.\,\ref{Fig:CMD_bck_KBD} shows the location of the predicted background galaxies in the $K/(J-K)$ CMD, without any reddening by the Orion nebula cloud. The intrinsic colors of the galaxies suffer from a redshift-dependent
  K-correction, i.e. they display redder $J-K$ colors compared to the background stars. The galaxies fainter than $K=18.5\,$mag affect only the isolated planetary mass objects below $0.006\,\unit{M_{\sun}}$, where our $J-K=2.6$ cut leads to incompleteness. Therefore we here focus on the galaxies brighter than $K=18.5\,$mag. 
  Adding about $2\,$mag ($8\,$mag) of extinction, the reddest (bluest) galaxies are shifted beyond our $J-K=2.6$ line. The background galaxies brighter than $K=18.5\,$mag have a median $J-K=1.6$. About $6\,$mag of
  extinction shifts such a galaxy out of our sample. As a conservative estimate for galaxies contributing as background contamination, we adopt a reddening of the cloud with less than $A_V=7\,$mag. The area of the HAWK-I mosaic with $A_V<7\,$mag inferred from the CO maps is $0.02$ square degrees. We scaled the number of predicted background galaxies to this area ($N_{galaxies}$ $\sim$670). 
  
  The results from other CMDs are similar. For CMDs using the $H$-filter, we adopt intrinsic colors $H-K=0.5\cdot (J-K)$; we find that the resulting IMFs are similar for $H-K=0.4 \cdot (J-K)$ and $H-K=0.6 \cdot (J-K)$ as well, hence not very sensitive to the assumption on the galaxies' colors.   
  
  As for the background stars, each galaxy is also moved along the reddening vector and the magnitude of the intersection is assigned to it. In Sec.\,\ref{decontamination_of_the_LF} we subtract the contamination caused by the galaxies and the background stars from the ONC data in order to find the LF of ONC members.
  \begin{figure}
      \centering
      \includegraphics[width=\hsize,keepaspectratio, angle=0, trim=0cm 0cm 0cm 0cm, clip=true]{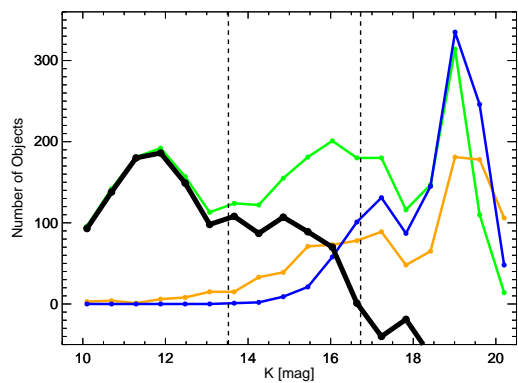}
        \caption{K-band luminosity function for the total \mbox{HAWK-I} data is shown in green, for background stars in yellow, and background galaxies in blue. Data and background objects are shifted to the isochrone along the $A_V$ vector. The black line shows the total HAWK-I data minus background stars + galaxies. The vertical dash-dotted lines mark the luminosity boundaries of brown dwarfs.}  
        \label{fig:KLF_all}
  \end{figure}
  
\section{Decontaminated luminosity function and comparison with the literature} \label{decontamination_of_the_LF}
   Starting from the LF of all sources in the field we first subtract the LF for reddened stars and then subtract the LF for reddened galaxies. Fig.\,\ref{fig:KLF_all} displays the LF of both total observed population and predicted background sources shifted (dereddened) along the direction of the extinction vector to the 3\,Myr isochrone. Note that this dereddening changes the resulting LFs, because highly reddened source now become brighter. The black curve shows the decontaminated LF. It exhibits a remarkably clear dip at the BD limit and a peak in the BD range at $K\,\sim\,16$\,mag. Beyond the BD range ($K>17.5$\,mag) the LF becomes jagged, probably caused by features of the isochrone (almost parallel to the $A_V$ vector). 
    
   Beside stellar maximum as well known from \cite{Muench02} and \cite{Robberto2010} (Fig.\,\ref{Comparison_Rob}) there is a second rise into the substellar regime peaking at $K=16$\,mag.   
 
\begin{figure}
     \centering
     \includegraphics[width=\hsize,keepaspectratio, angle=0, trim=0cm 0cm 0cm 0cm, clip=true] {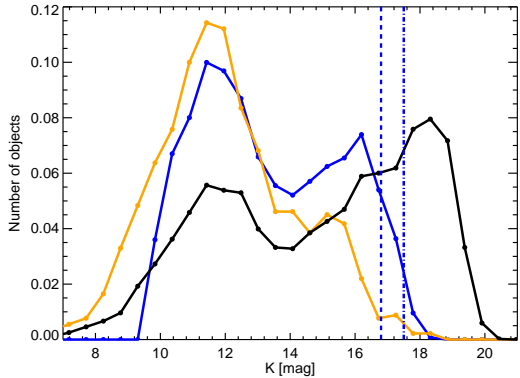}
     \caption{$K$-band luminosity functions normalized to the total number of sources in each sample. The HAWK-I data are marked in black and compared with \citet{Robberto2010} (blue) and \citet{Muench02} (yellow).  For all samples the original values are shown. No de-reddening, i.e. no shifting to the isochrone, no completeness correction and no foreground or background subtraction was done. The vertical dashed and dash dotted lines mark the completeness limits of \citet{Robberto2010} at 90\% and 70\%, respectively.}
     \label{Fig:KLF_norm}
   \end{figure}

   The surveyed areas of Robberto's, Muench's and our analysis are very different (see, Fig.\,\ref{Fig:FoVs}). 
   To compare the relative differences between the KLF -- features the LF normalized to the total number of sources in each sample is presented in Fig.\,\ref{Fig:KLF_norm}. The feature in the substellar range is about $30\%$ higher than the stellar maximum for the HAWK-I data, while for Robberto and Muench the ratio is vice versa. The maximum in the stellar mass range has about $25\%$ more sources for Robberto and even $\sim 250\%$ for Muench compared to the feature in the substellar range. This presents an additional hint to the incompleteness that Robberto's and Muench's analysis are suffering from. Therefore, both analysis can not present the same results as found here in this analysis. 
   From the direct comparison between the data from \cite{Robberto2010} and the HAWK-I data set, the 90\% completeness limit for the \cite{Robberto2010} is at $\sim\,17$\,mag (Fig.\,\ref{Comparison_Rob}, dashed line). At the Brown Dwarf limit ($\sim\,18$\,mag) Robberto presents a completeness 70\% ($\sim$\,120 sources). Using the number counts determined with HAWK-I (Fig. \ref{Comparison_Rob}, dashed dotted line) $\sim\,370$ sources are detected. Hence the completeness at $K=18$\,mag is only about 30\%. Robberto's estimation was done by recovering artificially introduced stars. These might be easier to recover than the actual object and therefore yield a too high completeness.
   
\begin{figure}
     \centering
     \includegraphics[width=\hsize,keepaspectratio, angle=0, trim=0cm 0cm 0cm 0cm, clip=true] {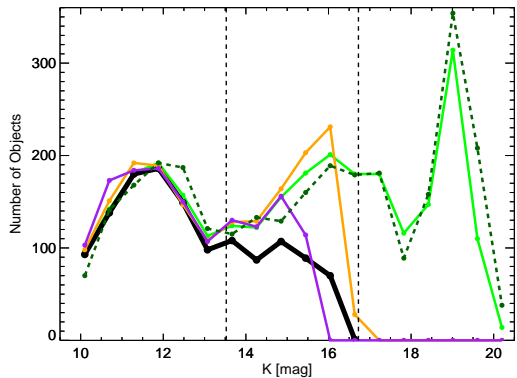}
     \caption{K-band luminosity function comparisons. The total \mbox{HAWK-I} data, shifted to the isochrone, are shown in green. The black line shows the total, shifted HAWK-I data minus background stars + galaxies using the latest model from \citet{Baraffe15} and \citet{Allard13}. The dark green dashed line presents the data shifted in K by $0.2$\,mag to account for a possible $K$-band excess. The orange line shows the KLF using the Isochrone from \citet{DAntona98} and the purple line presents the KLF using the Isochrone from \citet{Baraffe98}.}
     \label{Fig:Comparison_Dantona_Baraffe_K-execess}
   \end{figure}

   To test the consequences of a possible $K$-band excess on the KLF, in Fig.\,\ref{Fig:Comparison_Dantona_Baraffe_K-execess} the dashed green line shows the KLF of all our sources shifted by $+0.2$ in $(H-K)$ (average excess from \citet{Muench02}). This probably overestimates the apparent $K$-band excess, since also all potential background sources are shifted. Nevertheless, the KLF appears shifted by about 0.5\,mag to fainter magnitudes but maintains essentially the same feature shape.
   
   To analyse the influence of the isochrone used here, we compare older isochrones from \citet{Baraffe98} and \citet{DAntona98} as used in \cite{DaRio11}. The result is presented in Fig.\,\ref{Fig:Comparison_Dantona_Baraffe_K-execess}. These shape is very similar but the older isochrones do not cover the complete BD range and are therefore not considered in the following analysis.

\section{From the luminosity function to the IMF}
  In order to compute the IMF of the ONC, we subtract the background contamination in the mass space.
  
  Mathematically the subtraction of the ``\,mimicked\,'' background mass function (MF) is equivalent to constructing the LFs and subtracting the background contamination from the LF and then bin the mass function. However, the second procedure requires twice binning steps, one for the LF and one for the MF. The advantage of subtraction of the mimicked background mass function is that binning is required only once and that shot noise from the double binning is reduced.

\begin{figure}
   \centering
   \includegraphics[width=\hsize, keepaspectratio, angle=0, trim=0cm 0cm 0cm 0cm, clip=true] {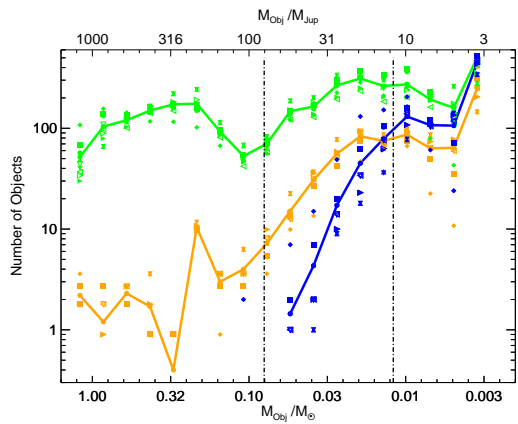}
   \caption{
     Mass functions. The symbols correspond to a single
     color-magnitude diagram, and the thick lines show the
     average. Total HAWK-I data (green), background stars (yellow),
     and background galaxies (blue). The vertical dash-dotted lines
     mark the mass boundaries of brown dwarfs. 
   }   
   \label{SMF_background}
  \end{figure}

  Together with the luminosity each potential (by potential we mean every detection) ONC member objects a mass was assigned according to the used isochrone. This procedure was applied to the nine CMD combinations, eventually deriving nine mass distributions. These distributions have been converted into individual IMFs by counting the number of sources in bins of 0.56\,dex in log(M) space. Finally we constructed an average IMF. Despite some redundancy of the CMDs, dispersion of the individual IMFs allows to get a view of the uncertainties relative to deriving masses with the same models but different data. Using the entire extinction limited ($A_V = 10$\,mag) HAWK-I data set yields the ``\,total IMF\,'' as shown in Fig.\,\ref{SMF_background}.
  
  The ``\,total IMF\,'' contains not only ONC members but also background objects which are reddened by the ONC dust screen.  While assigning luminosity to the background stars and galaxies as described in Sec.\,\ref{Sec:Accounting_for_bck_obj} also a mass corresponding to the used isochrone was recorded for each object. 
  When creating the ``\,total IMF\,'' from the data, such background objects are erroneously classified as ONC members. To get rid of the background contamination in the IMF, we determine the ``\,mimicked\,'' background IMF
  from the predicted background sources shown in Fig.\,\ref{Fig:CMD_bck_KBD} by binning them in the same way as the data and then subtract them from the ``\,total IMF\,''. Note that our aim is not to reconstruct the true type or mass of the background objects, rather we need to calculate their contribution to ONC's IMF. We created the ``\,mimicked\,'' background IMF for the background stars and galaxies, for each of the nine CMD combinations as we did for the total IMF. For both the background stars and the galaxies the resulting ``\,mimicked\,'' IMFs shown in Fig.\,\ref{SMF_background} are steeply rising with decreasing mass and remarkably smooth. In the same way as for the member LF for the ONC, the subtraction of the ``\,mimicked\,'' background IMFs from the total IMF results in the ONC member IMF that is shown in Fig.\,\ref{SMF}. Summarizing our approach to deal with contaminants, we shift the simulated populations (both background stars and galaxies) along the $A_V$ vector to the isochrone to treat them in the same way as the objects in the cloud.
  
  In order to take into account propagation of the photometric errors in the mass estimations, we run a Monte-Carlo Simulation over a member candidate population determined by making extended use of the extinction maps based on \citet{Shimajiri2011} and \citet{Ripple13} (paper in preparation).
  
  The total extinction caused by the cloud can be used to distinguish foreground stars from objects embedded in the star forming region or from background objects. For each object we compare the total extinction - as derived from CO measurements - with the extinction along the line of sight to individual objects - as derived from the color-magnitude diagram. Objects behind the cloud suffer the same or even higher extinction than what is obtained from CO measurements. Objects within the cloud and foreground stars should display lower extinction values than the total extinction throughout the cloud. This procedure assumes that there does not exist a significant contribution of circumstellar extinction due to an opaque disk which would probably remain undetected by the CO measurements. Foreground objects are discarded using the analysis from \citet{Alves12} and \citet{Bouy14}. This method is limited by the lower resolution of the CO extinction maps and might therefore result in slightly different number counts.
  
  For each data point we obtaining $100$ realizations.This provided us with individual masses and uncertainties for every isochrone used. We used four different isochrones: $1, 2, 3$ and $5$\,Myrs respectively to asses the effect of the age in the derived mass function, and finally, to be phase independent in the representation of such mass function, we estimated the Kernel Density Estimator of each mass distribution (for each isochrone, where we are normalizing to make them directly comparable and also with ONC member mass distribution of Fig.\,\ref{SMF}), with its 99\% confidence level. The results  using the Monte-Carlo Simulation together with the Kernel Density Estimator are shown in Fig.\,\ref{Fig:SMF_age_comparsion}. In comparison with counting the number of objects in each mass bin the results are similar.
  
\section{Resulting Initial Mass Functions}
  \begin{figure}
   \centering
   \includegraphics[width=\hsize, keepaspectratio, angle=0, trim=0cm 0cm 0cm 0cm, clip=true] {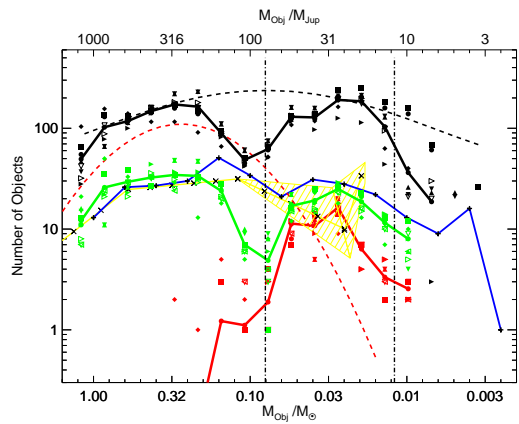}
   \caption{ 
     Initial mass functions. The symbols (open and filled) correspond to a single color-magnitude diagram, and the thick solid lines show the average. Members of the entire Orion Nebula Cloud (black) were calculated from Fig. \ref{SMF_background} as the difference between the total IMF and the IMF of background stars and galaxies. The green solid line gives the mass function of a small subarea containing the known BDs South and West of the Trapezium, where no background contamination is expected. The red solid line gives the mass function of the known spectroscopically verified BDs located in the subarea and shown as red dots in Fig.\,\ref{Fig:CMD_bck_KBD}. Vertical dash-dotted lines mark the mass boundaries of BDs. For comparison other IMFs are shown: black dashed line: \citet{Chabrier05} standard IMF; \citet{DaRio11} red dashed line which -- inconsistently -- falls below the IMF of the known BDs (red solid line);  
     \citet{Muench02} yellow solid line with thin black crosses and shaded area giving the error range, note the abrupt upturn in the last bin; \citet{Lucas05} blue solid line with thin black plus symbols.
     }
   \label{SMF}
  \end{figure}
  
  The striking features of the ONC member IMF (Fig.\,\ref{SMF}) is the presence of two peaks at about 0.25 and 0.025\,$\unit{M_{\sun}}$ separated by a pronounced dip at the hydrogen burning limit (0.08\,$\unit{M_{\sun}}$), which corresponds to the zone of low object density already seen in Fig.\,\ref{Fig:CMD_data}. The IMF extends into the mass regime of IPMOs ($<0.012\,\unit{M_{\sun}}$). The IMF contains 929 stars with $M<1.4\,\unit{M_{\sun}}$, 757 BD candidates and 158 IPMO candidates with $M\,>\,0.005\,\unit{M_{\sun}}$, hence indicates a high fraction ($\sim$50\%) of substellar objects, about ten times more than previously estimated.  
  
  We account for the uncertainty in the estimated masses by determining the masses from the different CMDs. The dispersion is displayed by the points in the IMF. For the Orion members it is minimal at $0.5\,\unit{M_{\sun}}$ with about $\pm 10\%$ and maximal at $0.07\,\unit{M_{\sun}}$ with about $+ 40\%/-30\%$.

  The result on the bimodal IMF may be sensitive to the number of background sources and the assumptions on the CO-derived extinction of the ONC. To entirely remove the substellar IMF peak one needs to increase the number of background objects by factor of two to three. It is not yet clear whether such a large increase could be present, for instance due to clumpiness of ONC's dust screen. To test this Hypothesis, we performed an independent check, using the known spectroscopically verified BDs (see red dots in Fig.\,\ref{Fig:CMD_bck_KBD}). These BDs are located in the $3\arcmin-6\arcmin$ wide stripe South and West of the Trapezium which has been observed with the Gemini-S telescope \citep{Lucas05}. This subarea shows a high CO column density, corresponding to $A_V>20\,$mag, hence it is very unlikely that in this area our extinction limited data set ($A_V<10\,$mag) is affected by background contamination. Thus without any background subtraction, we constructed the IMF of that complete subarea using the HAWK-I data, which show about 50\% more objects than the former Gemini-S data.  

  Fig.\,\ref{SMF} displays the new IMF of the subarea (green) and the IMF of the spectroscopically confirmed BDs therein (red). The subarea IMF has also a bimodal shape and peaks at about 0.25 and $0.025\,\unit{M_{\sun}}$ -- exactly as the IMF of the entire ONC area; additionally, compared to the confirmed BDs (red), there are about 50\% more BDs predicted (green). While for the IMF of the entire ONC area the substellar
  peak is equally high as the stellar peak, for the IMF of the subarea the height of the substellar peak is about 30\% lower than the stellar peak; this may be due to the difficulty to detect faint sources against the nebula emission in the subarea. The check on the subarea gives further support for the bimodality of the IMF for the entire $22\arcmin \times 28\arcmin$ ONC area mapped with \mbox{HAWK-I}\footnote{ Recently, we have obtained KMOS/VLT spectra of 20 new BD candidates, selected form the HAWK-I data, 16 of which turned out to be in fact BDs. They are located towards a region with a CO-screen of $A_V \sim 7\,$mag. The high BD fraction (80\%) indicates that the contamination of the IMF by background objects is low (Drass et al. in prep.)}.
  To make a further statistical test, we employ the widefield optical and NIR data set provided by \citet{Bouy14}. Cross-correlating both catalogs gives $918$ common objects within a $3"$ matching radius. Using all bands, for $664$ objects the membership probability given in \citet{Bouy14} is less than $90\%$ and for $649$ objects the probability is $<60\%$. Both numbers are consistent with $890$ background stars and $670$ galaxies found in this statistical analysis. In the BD mass range ($13.8\,$mag$\,<\,K\,<\,17.2\,$mag) \citet{Bouy14} present $150$ sources with a membership probability of less than $60\%$. Consistently, we find more background sources, namely $192$ galaxies and $309$ background stars.

\begin{figure}
   \centering
   \includegraphics[width=\hsize, keepaspectratio, angle=0, trim=1.5cm 1cm 3cm 1cm, clip=true] {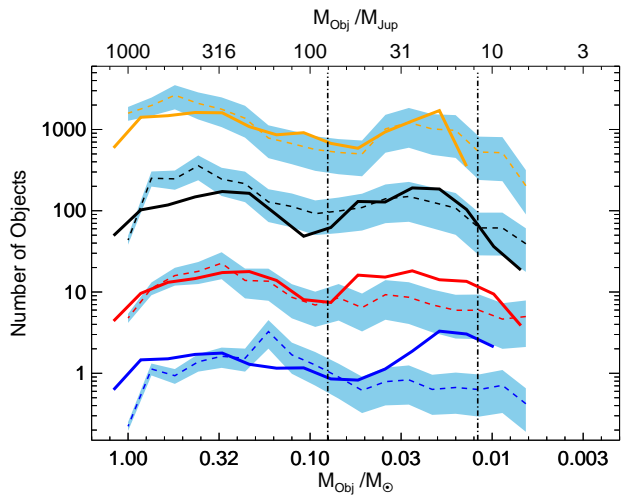}
   \caption{{HAWK-I Initial mass functions using the $1,2,3$ and $5$\,Myr isochrone from \citet{Baraffe15} and \citet{Allard13} are shown by the solid blue, red, black, and orange lines, respectively. The dashed lines show the results from the Monte-Carlo Simulation using a Kernel Density Estimator. For easier identification, except for the 3\,Myr isochrones, the pairs of isochrones are offset by factors of ten. Confidence intervals of 99\% are displayed as blue shaded areas.} The different colors represent the same ages as for the solid lines. The marks are the same as in Fig\,\ref{SMF}.}  
   \label{Fig:SMF_age_comparsion}
  \end{figure}

  For comparison, Fig.\,\ref{Fig:SMF_age_comparsion} shows the IMF assuming an age of $1, 2, 3$ and $5$\,Myrs for the ONC. The position of the peak in the BD mass regime using the $2, 3$ and $5$\,Myr isochrone is mainly unchanged and the minimum between the peak in the stellar mass range and in the BD mass range appears to be wider. The results for the 1\,Myr isochrone shows the BD peak at lower masses ($\approx 20\,M_{Jup}$). Nevertheless, the overall shape is essentially preserved.
  
  For the Monte-Carlo Simulation it can be seen in Fig.\,\ref{Fig:SMF_age_comparsion}, while the double peak structure is not present in the case of the 1\,Myr isochrone, it is clearly insensitive to the age use in the range $2-5$\,Myr. We interpret this fact pointing out the commonly assumed issue that younger isochrones are more subject to unaccounted processes. And that in fact, the double peak feature is a strong characteristic of this distribution.
  
  From the direct comparison between the KDE of the Monte-Carlo simulation and the simple number counts, it can be concluded that both methods agree within the uncertainties for ages $>\,2$\,Myr. The overabundance of sources at 1\,Myr from the simple count approach may reflect its lack of robustness against small variations, especially at the lower end of the mass spectrum, where the theoretical isochrones run nearly horizontally in the CMD. 
  
  In the following we compare our IMFs with previous results. The ONC IMF by \citet{DaRio11} falls clearly below the IMF of known BDs (compare the red curves in Fig.\,\ref{SMF}). Because the spectroscopically confirmed BDs are contained in the $\sim$30$\arcmin$$\times$30$\arcmin$ area mapped by \citet{Robberto2010} and used by \citet{DaRio11}, their IMF appears puzzling. It may be affected by incompleteness and overcorrection of background contamination, and the 4-m telescope ISPI data may be too shallow and/or offering too low angular resolution for detecting the BDs in the Gemini-S subarea. A similar decline of the inner ONC IMF across the Hydrogen burning limit has also been reported by (\citet{Hillenbrand00} their Fig. 16\,) based on their quite shallow Keck observations of the central 5$\arcmin$$\times$5$\arcmin$ area, which shows very bright nebula emission. Compared to the log-normal distribution parametrization of \citet{Chabrier05} IMF (black dashed curve in  Fig.\,\ref{SMF}), the depth of the dip at 0.08\,$\unit{M_{\sun}}$ in our new ONC IMF is about a factor of 3 (for the used mass bin sizes). \citet{Muench02} have reported a decline of the ONC IMF with a potential upturn at 0.02\,$\unit{M_{\sun}}$, but -- in view of their small field (5$\arcmin$$\times$5$\arcmin$), limited depth of the data and the large error bars -- the claim of an IMF dip and low-mass upturn based on the last mass bin only appears speculative (yellow curve in Fig.\,\ref{SMF}). 
  The ONC subarea IMF by \citet{Lucas05} (blue curve in Fig.\,\ref{SMF}) has been calculated with larger mass bins, which lead to higher numbers per bin and tend to smear out details. Nevertheless, accounting for the smaller number of sources in the Gemini-S data ($\sim$400 vs $\sim$600 in the VLT data), Lucas et al.'s IMF is nicely consistent with our more detailed bimodal IMF from the VLT HAWK-I data (green solid curve). While the previous IMFs have been derived using older isochrones, the isochrone differences at the mass range above 0.03 $M_\odot$ are too small to explain any of the coarse IMF differences seen.

  \section{On the origin of the bimodal IMF}

  The explanation for the bimodal shape of the IMF is not straight forward. The IMF dip occurs just at the Hydrogen burning limit, suggesting at a first glance that the burning mechanism of young stars and BDs may play a role. However, the young BD objects gain still most of their energy from mass accretion and  gravitational contraction, so that the Hydrogen burning plays a minor role. Therefore other mechanisms might be needed to explain the origin of the bimodal IMF. 

  The predicted starless core mass function (CMF) for Orion, using an $A_V$-dependent extrapolation method, steadily declines between the peak at 1\,$\unit{M_{\sun}}$ and 0.1\,$\unit{M_{\sun}}$, and at lower mass it stays on a constant plateau at a level of a factor 5 lower than the peak height \citep{Sadavoy10}. This CMF shape is hardly consistent with the bimodal IMF. It challenges deterministic theories which emphasize the role of a direct mapping between CMF and IMF \citep{Nutter07, Hennebelle13, Andre10}. Therefore an attractive explanation for the bimodal IMF could be ejection of low mass objects from small groups of protostars \citep{Reipurth01} or from fragmenting circumstellar disks \citep{Vorobyov13}.   

  While simulations of fragmenting gas clouds indicate that BDs can be produced by the ejection mechanism, the resulting stellar/substellar IMFs often show only a modest (factor $<$ 2) and smoothly running excess of substellar objects above the log normal extrapolation from the stellar IMF peak (e.g. \citet{Bate09}, his Figs. 7 \& 8). So far, many simulations yield a continuous mass spectrum of the ejected objects, which results in a broad featureless IMF. This contradicts the structured IMF and the distinct substellar peak that we observed for the ONC. On the other hand, recent simulations of a fragmenting circumstellar disk of a 1.2\,$\unit{M_{\sun}}$ pre-stellar core indicate that the ejected fragments have a mass spectrum of BDs and IPMOs (\citet{Vorobyov13}, their Fig. 4a). We conclude that the CMF might provide a basis of objects for the later IMF but the excess of BDs and IPMOs is likely produced by ejected fragments from circumstellar disks. We suggest that the final explanation for Orion's bimodal IMF has to be searched for along that direction.

  \section*{Acknowledgments}
   The author would like to thank their anonymous referee for her/his helpful comments.
   HD likes to thank ESO, Chile for the generous support during a two year studentship. 
   HD acknowledges financial support from FONDECYT project 3150314.
   A.\,Bayo acknowledges financial support from the Proyecto Fondecyt Iniciaci\'on 11140572 .
   This work was supported by \emph{Deut\-sche For\-schungs\-ge\-mein\-schaft, DFG\/} project number Ts~17/2--1, 
   as well as by the Nordrhein-Westf\"alische Akademie der Wissenschaften und der K\"unste  in the framework of the academy program of the Federal Republic of Germany and the state Nordrhein-Westfalen.

\appendix
 \section{Catalog of ONC members}\label{app_member_cat}
 An example for the HAWK-I source catalog is shown in \mbox{Table\,\ref{member_cat}}. The full catalog is available online.
       \begin{table*}
         \caption{Catalog of the HAWK-I sources}
         \label{member_cat}
         \centering
         \begin{tabular}{ccccccccccc}
           \hline\hline   \\
           RA (J2000) &  Dec (J2000) &   J [mag] & uncJ [mag] & H [mag]  & uncH [mag] & K [mag]  & K [mag]\\
           \hline\\
           83.63357544 & -5.17309999  & 15.299  &  0.009   & 14.813 &  0.013   & 14.456 & 0.013    \\
           83.63535309 & -5.34161282  & 14.177  &  0.010   & 13.263 &  0.010   & 12.857 & 0.022    \\
           83.63700867 & -5.22061920  & 17.710  &  0.010   & 16.581 &  0.015   & 16.398 & 0.025    \\
           83.63703156 & -5.17097521  & 16.234  &  0.008   & 15.524 &  0.014   & 14.911 & 0.009    \\
           83.63822174 & -5.16701651  & 18.278  &  0.009   & 17.425 &  0.019   & 16.673 & 0.017    \\
           83.63856506 & -5.12406111  & 14.863  &  0.001   & 14.177 &  0.001   & 13.930 & 0.001    \\
           83.63997650 & -5.13867235  & 17.839  &  0.004   & 16.955 &  0.007   & 16.624 & 0.011    \\
           83.64015961 & -5.20838356  & 20.423  &  0.039   & 19.270 &  0.023   & 18.884 & 0.050    \\
           \hline
         \end{tabular}
       \end{table*}
   
 \section{Isochrone}\label{App:Allard_3Myr_isochrone}
 The 3\,Myr isochrone from F. Allard (download on 25.05.2015) is presented in Table \ref{Table:Allard_3Myr_isochrone}.
        \begin{table*}
         \caption{3\,Myr isochrone from F. Allard.}
         \label{Table:Allard_3Myr_isochrone}
         \centering
         \begin{tabular}{ccccccccccc}
           \hline\hline   \\
            M/Ms  & Teff(K)  &  L/Ls & lg(g) & R(Gcm)  &     D   &   Li  &         J    &       H   &    K    \\
            \hline\\
            0.0005 &  501. &  -5.87 &  2.75 & 10.86 & 1.0000 & 1.0000  &    19.202  &    19.296   &   19.037  \\
            0.0010 &  734. &  -5.26 &  3.10 & 10.23 & 1.0000 & 1.0000  &    15.897  &    15.831   &   15.694  \\
            0.0020 & 1025. &  -4.65 &  3.38 & 10.49 & 1.0000 & 1.0000  &    14.498  &    13.517   &   13.392  \\
            0.0030 & 1241. &  -4.29 &  3.52 & 10.93 & 1.0000 & 1.0000  &    13.873  &    12.602   &   12.035  \\
            0.0040 & 1415. &  -4.03 &  3.62 & 11.34 & 1.0000 & 1.0000  &    13.418  &    12.186   &   11.306  \\
            0.0050 & 1580. &  -3.80 &  3.68 & 11.79 & 1.0000 & 1.0000  &    13.180  &    11.836   &   10.796  \\
            0.0060 & 1721. &  -3.62 &  3.72 & 12.27 & 1.0000 & 1.0000  &    12.187  &    11.123   &   10.322  \\
            0.0070 & 1842. &  -3.46 &  3.76 & 12.77 & 1.0000 & 1.0000  &    11.599  &    10.686   &   10.024  \\
            0.0080 & 1943. &  -3.34 &  3.78 & 13.29 & 1.0000 & 1.0000  &    11.075  &    10.320   &    9.760  \\
            0.0090 & 2028. &  -3.23 &  3.79 & 13.86 & 1.0000 & 1.0000  &    10.674  &    10.035   &    9.538  \\
           
           \hline
         \end{tabular}
       \end{table*}
   
\bsp
\label{lastpage}

\end{document}

%% file: library_abkuerzungen.tex
\def\aj{AJ}%
\def\araa{ARA\&A}%
\def\apj{ApJ}%
\def\apjl{ApJ}%
\def\apjs{ApJS}%
\def\ao{Appl.~Opt.}%
\def\aap{A\&A}%
\def\mnras{MNRAS}%
\def\pasj{PASJ}%